\documentclass[12pt]{article}
\renewcommand{\theequation}{\thesection\arabic{equation}}

\def\vep{\varepsilon}

\def\beq{\begin{equation}}
\def\eeq{\end{equation}}

\def\bea{\begin{eqnarray}}
\def\eea{\end{eqnarray}}

\begin{document}

\begin{center}
{\huge
{\bf Gauge Models\\ in Modified Triplectic Quantization\\}}

\vspace{1cm}

{\large
{\sc B.~Geyer}$^{\ a,b,}$ \footnote{E-mail: geyer@itp.uni-leipzig.de},
{\sc P.M.~Lavrov}$^{\ a,c,}$ \footnote{E-mail: lavrov@tspu.edu.ru} and
{\sc P.Yu.~Moshin}$^{\ c}$}\\

\vspace{1cm}

$^a)$ {\normalsize\it Center of Theoretical Studies,
Leipzig University,\\
Augustusplatz 10/11, 04109 Leipzig, Germany}

\vspace{.5cm}

$^b)$ {\normalsize \it Instituto de Fisica, 
Universidade de S$\tilde a$o Paulo,\\
Caixa Postal 66318-CEP, 05315-970 S$\tilde a$o Paulo, SP, Brazil}

\vspace{.5cm}

$^c)$ {\normalsize\it Tomsk State Pedagogical
University, 634041 Tomsk, Russia}\\
\end{center}

\vspace{.5cm}

\begin{quotation}
\noindent
\normalsize
 The modified triplectic quantization is applied to several
 well-known gauge models: the Freedman--Townsend model of non-abelian
 antisymmetric tensor fields, $W_2-$gravity, and  2D gravity with
 dynamical torsion. For these models we obtain explicit solutions of
 those generating equations that determine the quantum action and the
 gauge-fixing functional. Using these solutions, we construct the
 vacuum functional, determine the $Sp(2)$-invariant effective actions and
 obtain the corresponding transformations of extended BRST symmetry.
\end{quotation}

\section{Introduction}

 In  recent  years the development of covariant quantization rules for
 general  gauge  theories  on  the basis of extended BRST symmetry has
 become increasingly popular \cite{Sp2} --
\cite{TT}.

 The realization of the principle of extended BRST symmetry, combining
 BRST  \cite{BRST} and anti-BRST \cite{antiBRST} transformations,
 naturally  unifies the treatment of auxiliary
 variables  that  serve  to  parametrize  the  gauge in the functional
 integral  and  those  that enter the quantum action determined by the
 corresponding  generating  equations.  Basically,  the above tendency
 manifests  itself in enlarging the configuration space of the quantum
 action    with   auxiliary   gauge-fixing   variables   
 (see,~e.g.,~Refs.~\cite{Sp2,BM,L}).  
 Recently,  however,  it  has  been strengthened by
 extending  the  concept  of  generating  equations to cover the case
 of introducing the gauge \cite{BM,GGL}.

 The  method  of  $Sp(2)$-covariant quantization \cite{Sp2} was one of
 the  first  to  provide  a  realization of extended BRST symmetry for
 general  gauge theories, i.e., theories of any stage of reducibility with
 a  closed  or  open  algebra  of  gauge transformations. The complete
 configuration  space  $\phi^A$  of a gauge theory, considered in this
 approach,  is  constructed  by  the  rules  of  the  BV  quantization
 \cite{BV}  and  consists of the initial classical fields supplemented
 by  the  pyramids of auxiliary variables, i.e., ghosts, antighosts and
 Lagrange   multipliers,  according  to  the  corresponding  stage  of
 reducibility.  Even  though  these auxiliary variables originally
 \cite{BV}  play  different  roles  in the construction of the quantum
 theory,  their  consideration  within the $Sp(2)$-covariant formalism
 allows  to achieve a remarkable uniformity of description. Namely, in
 the  framework  of  the  $Sp(2)$-covariant approach, the pyramids of
 ghosts are combined with the corresponding pyramids of antighosts and
 Lagrange  multipliers into irreducible representations of the group
 $Sp(2)$,  which form completely symmetric $Sp(2)$-tensors and enter the
 quantum  theory  on equal footing in terms of both the quantum action
 and  the gauge fixing functional.  The quantum action of the 
 $Sp(2)$-covariant
 formalism  depends  on  an  extended  set  of  variables,
 including,  besides  the fields $\phi^A$, also the sets of antifields
 $\phi^*_{Aa}$  and  $\bar{\phi}_A$. In the case of linear
 dependence of the quantum action on
 $\phi^*_{Aa}$  and  $\bar{\phi}_A$
 they may be interpreted as sources  of extended BRST transformations
 and sources of mixed BRST and anti-BRST transformations,
 respectively.

 In  \cite{L}, a consistent superfield formulation of the 
 $Sp(2)$-covariant rules was  proposed.  This  approach allows to combine 
 all the variables of
 the  $Sp(2)$-covariant  formalism,  namely, the fields and antifields
 $(\phi^A,\phi^*_{Aa},\bar{\phi}_A)$  that  enter  the quantum action,
 the auxiliary variables $(\pi^{Aa},\lambda^A)$  that  serve  to
 parametrize the gauge, and, finally, the sources $J_A$ to the fields
 $\phi^A$,   into  superfields $\Phi^A(\theta) = \phi^A + 
 \pi^{Aa}\theta_a +
 \frac{1}{2}\lambda^A\theta_a\theta^a$ and  superantifields (supersources)
 $\bar{\Phi}_A(\theta) = {\bar\phi}_A - \theta^a\phi^*_{Aa}- \frac{1}{2}
 \theta_a\theta^a J_A$, defined  on  a  superspace  with two scalar Grassmann
 coordinates $\theta_a$. The quantum action of this theory is defined as a
 functional of superfields and superantifields, $S(\Phi,{\bar\Phi})$,
 which makes it possible  to  realize the transformations of extended BRST
 symmetry in terms of  supertranslations along the Grassmann coordinates.

 Moreover, in the recent paper \cite{GM99} the superspace approach 
 was extended
 by considering not only the (sub)group of translations but also 
 the full group
 of conformal transformations on the superspace with two Grassmann 
 coordinates.
 The generators of this conformal group span an algebra isomorphic to the
 superalgebra $sl(1,2)$. In this approach it is possible to consider massive
 gauge theories by introducing mass-dependent BRST and antiBRST operators
 that
 are related to translations coupled (with the factor $m^2$) to special 
 conformal
 transformations. Furthermore, the transformations of $Sp(2)$-symmetry, 
 including
 the symmetries which underly the conservation of the ghost number and  
 the "new
 ghost number", are realized as (symplectic) rotations and dilatations, 
 respectively.

 In the framework of the triplectic quantization \cite{BM}, another 
 modification
 of  the $Sp(2)$ covariant approach was proposed, based on a different
 extension  of  the configuration space of the quantum action. Namely,
 it  was  suggested  to  consider  the  auxiliary fields $\pi^{Aa}$ as
 variables anticanonically conjugated to the antifields $\bar{\phi}_A$
 with   the  corresponding  redefinition  of  the  extended
 antibrackets  \cite{Sp2} which appear in the generating equations for
 the  quantum  action.  Another  feature  of  the triplectic formalism
 is  that  the  gauge-fixing  part  of  the  action  in  the
 functional  integral  is  determined by generating equations formally
 similar to the equations that describe the quantum action. The entire
 set  of   variables  necessary for the construction of the vacuum
 functional   in   the   triplectic   formalism   coincides  with  the
 corresponding  set  of the $Sp(2)$-covariant approach and is composed
 by  the  fields  $(\phi^A,\phi^*_{Aa})$ and $(\pi^{Aa},\bar{\phi}_A)$
 anticanonically conjugated to each other in the sense of modified
 antibrackets, as  well  as by the remaining auxiliary fields $\lambda^A$
 that serve  to parametrize the gauge-fixing functional.

 In  the  recent  paper  \cite{GGL},  a  modification of the triplectic
 formalism was proposed, whose essential ingredients we will now briefly 
 review.

 While retaining the space of variables of the triplectic formalism and
 accepting the idea of imposing generating equations on both the quantum 
 action
 and the gauge-fixing functional, the formalism \cite{GGL} modifies the system
 of these equations, as well as the definition of the vacuum functional, 
 in order
 to ensure the {\it correct boundary condition} for the quantum action,
\[
 W\big|_{\Phi^* = \bar\phi = \hbar=0}= S_{\rm cl},
\]
 which explicitly allows to take into account the 
 information contained in the
 classical action. It also implies that the classical action of a theory
 satisfies  (in the limit $\hbar\rightarrow 0$) the  generating  
 equations for the quantum action
 $W=W(\phi,\phi^*,\pi,{\bar\phi})$,
\begin{equation}
\label{S}
\frac{1}{2} (W,W)^a + V^a W = i\hbar\Delta^a W,
\end{equation}
 in complete analogy with earlier quantization schemes, and in
 contrast to the original triplectic formalism \cite{BM}. The gauge-fixing
 functional $X=X(\phi,\phi^*,\pi,{\bar\phi};\lambda)$
 of the modified triplectic formalism satisfies similar
 generating equations,
\begin{equation}
\label{X}
 \frac{1}{2} (X,X)^a - U^a X = i\hbar\Delta^a X.
\end{equation}
 The above generating equations are expressed in terms of
 the extended antibrackets 
 \begin{eqnarray}
\nonumber
(F,G)^a &=& \left( \frac{\delta F}{\delta\phi^A}
                 \frac{\delta G}{\delta\phi^*_{Aa}}+
 \varepsilon^{ab}\frac{\delta F}{\delta\pi^{Aa}}
                 \frac{\delta G}{\delta{\bar\phi}_A}\right)
  - (F \leftrightarrow G) (-1)^{(\varepsilon(F)+1)(\varepsilon(G)+1)}\\
\nonumber
\end{eqnarray}
{and the differential operators}
\begin{eqnarray}
 \Delta^a &=& (-1)^{\varepsilon_A}
 \frac{\delta_l}{\delta\phi^A}\frac{\delta}{\delta\phi^*_{Aa}}
 +
 (-1)^{\varepsilon_A+1}\varepsilon^{ab}
 \frac{\delta_l}{\delta\pi^{Ab}}\frac{\delta}{\delta\bar\phi_A},
\\
\nonumber
 V^a &=& \varepsilon^{ab}\phi^*_{Ab}
 \frac{\delta}{\delta\bar\phi_{A}},
 \qquad
 U^a = (-1)^{\varepsilon_A+1} \pi^{Aa} \frac{\delta_l}{\delta\phi^A},
\end{eqnarray}
  where the derivatives with respect to the antifields are taken from
 the left, and $\varepsilon^{ab}$ is the antisymmetric tensor with the
 normalization $\varepsilon^{12}=1$. The operators $V^a$ and $U^a$
 are closely related to the differential operators which were introduced
 earlier in the framework of the superfield formalism \cite{L}, and which
 have a clear geometrical meaning as generators of supertranslations in
 superspace.

 Given the quantum action $W$ and the gauge-fixing functional $X$, the
 vacuum functional $Z\equiv Z(J=0)$ in the framework of the modified 
 triplectic
 quantization \cite{GGL} is defined by
\begin{eqnarray}
\label{Z}
 Z=\int d\phi\,d\phi^*d\pi\,d\bar{\phi}\,d\lambda\exp\left\{
 \frac{i}{\hbar}\left(W+X+S_0\right)\right\},\quad
 S_0=\phi^*_{Aa}\pi^{Aa}.
\end{eqnarray}
 Let us note that choosing the gauge fixing functional,
 $X=X(\phi,\pi,{\bar\phi};\lambda)$, in the form
\begin{eqnarray}
\label{F}
 X = \left( \bar{\phi}_A - \frac{\delta F}{\delta\phi^A}\right) \lambda^A
 -\frac{1}{2}\epsilon_{ab}U^a U^b F,
 \qquad F = F(\phi),
\end{eqnarray}
 solves eqs.~(\ref{X}) (with $\Delta^aX$ being identical zero).
 Then, the integrand in eq.~(\ref{Z}) is invariant under the following
 transformations (cf.,~Ref.~\cite{GGL})
\begin{eqnarray}
\label{exBRST}
\nonumber
 \delta\phi^{Aa}&=&-\left(\frac{\delta W}{\delta\phi^*_{Aa}}-
 \pi^{Aa}\right)\mu_a,
\\
\nonumber
 \delta\phi^*_{Aa}&=&\mu_a\left(\frac{\delta W}{\delta\phi^A}+
 \frac{\delta^2 F}{\delta\phi^A\delta\phi^B}\lambda^B+
 (-1^{\vep_A}\frac 12\pi^{Bb}
 \frac{\delta^3 F}{\delta\phi^A\delta\phi^B\delta\phi^C}
 \pi^{Cc}\right),
\\
 \delta\pi^{Aa}&=&\vep^{ab}\left(\frac{\delta W}{\delta{\bar\phi}_A}-
 \lambda^A\right)\mu_b,
\\
\nonumber
 \delta{\bar\phi}_A&=&\mu_a\vep^{ab}\left(\frac{\delta W}{\delta\pi^{Ab}}+
 \phi^*_{Ab}\right) +
 \mu_a\frac{\delta^2 F}{\delta\phi^A\delta\phi^B}\pi^{Ba},
\\
\nonumber
 \delta\lambda^A&=&0.
\end{eqnarray} 
 Here, $\mu_a$ is a doublet of constant anticommuting parameters.
 If, in addition,
 $W = W(\phi,\phi^*,{\bar\phi})$ is assumed not to depend on $\pi^{Aa}$ 
 then, obviously, the master equations (\ref{S}) and
 the vacuum functional (\ref{Z}) are reduced to those of the 
 $Sp(2)$-covariant formalism.

 The aim of this paper is to apply the above prescriptions of the modified
 triplectic formalism for quantizing several gauge models.

 In Section 2, we consider the model of an antisymmetric tensor field 
 suggested
 by Freedman and Townsend \cite{FT}. The Freedman--Townsend (FT) model is an
 abelian gauge theory of first stage reducibility. The corresponding complete
 configuration space is constructed by the rules of the $Sp(2)$-covariant
 formalism \cite{Sp2} for reducible gauge theories. In the case of the 
 FT model,
 the generating equations (\ref{S}) that determine the quantum action in the
 framework of the modified triplectic formalism can be solved exactly, which
 allows one to obtain the exact form of the vacuum functional in terms of the
 effective $Sp(2)$-invariant action $S_{\rm eff}$ as well as the
 corresponding transformations of extended BRST symmetry.

 In Sections 3 and 4, we consider the gauge model of $W_2-$gravity \cite{W2} 
 and
 the theory of two-dimensional gravity with dynamical torsion \cite{VK}, 
 respectively.
 Both these models are examples of irreducible gauge theories with 
 a closed algebra,
 and their configuration spaces are constructed by the rules of the 
 $Sp(2)$-covariant
 quantization for irreducible theories. 
 In order to obtain closed solutions of the above generating equations that
 determine the quantum action in the case of these gauge models, one has to
 introduce some regularization for $\Delta^a W$ when it occurs to be
 proportional to $\delta(0)$. Unfortunately, the regularizations which
 reduces all terms containing $\delta(0)$ to zero (see, e.g.,~\cite{Sp2}) 
 cannot be used here since both models under consideration are strictly
two-dimensional  and, therefore, it is not possible to use dimensional
regularization on which that procedure is based. Instead, one could  use
Pauli-Villars regularization as has been proposed in Ref.~\cite{Troost}
and applied in Refs.~\cite{JST} and \cite{VVP} to the case of $W_2$-- and
$W_3$--gravity, respectively. As an intermediate step we consider here only
the tree approximation, i.e., we determine explicit solutions of the classical
master equations. With them we obtain a closed form of the vacuum functional
and the corresponding transformations of extended BRST symmetry as well as
the related effective action, $S_{\rm eff}$, which depends on the fields only.

\section{Freedman--Townsend Model}
\setcounter{equation}{0}
 In the first order formalism, the theory of a non-abelian antisymmetric field
 $H_{\mu\nu}^p$, suggested by Freedman and Townsend \cite{FT}, is described
 by the  action \footnote{Contrary to the original notations \cite{FT}, 
 we denote the antisymmetric tensor field by $H$ 
 in order to avoid confusion with
 the auxiliary fields to be introduced below.}
\begin{equation}
\label{fts}
 S_{cl}(A_{\mu}^p, H_{\mu\nu}^p)=\int{d^4}x\left(-\frac{1}{4}
 {\varepsilon}^{\mu\nu\rho\sigma}F_{\mu\nu}^pH_{\rho\sigma}^p
 +\frac{1}{2} A_{\mu}^p A^{p\mu}\right),
\end{equation}
 where $A_{\mu}^p$ is an (auxiliary) gauge field with the strength
 $F_{\mu\nu}^p={\partial}_\mu A_{\nu}^p-{\partial}_\nu A_{\mu}^p+f^{pqr}
 A_{\mu}^qA_{\nu}^r$ (the coupling constant is absorbed into the
 structure coefficients $f^{pqr}$), and the Levi--Civita tensor
 ${\varepsilon}^{\mu\nu\rho\sigma}$ is normalized as ${\varepsilon}^{0123}=1$.
 Eliminating the auxiliary field $A^p_\mu$ with the help of the field equations
 leads to the more complicated action of the second order formalism \cite{FT}. 

 The action (\ref{fts}) is invariant under the gauge transformations
\begin{equation}
 \label{ftgauge}
 {\delta}A^p_{\mu}=0,\quad
 {\delta}H_{\mu\nu}^p={\cal D}_{\mu}^{pq}{\xi}_{\nu}^q -
 {\cal D}_{\nu}^{pq}{\xi}_{\mu}^q\equiv 
 {\cal R}_{\mu\nu\alpha}^{pq}{\xi}^{q\alpha},
\end{equation}
 where ${\xi}_{\mu}^p$ are arbitrary parameters, and ${\cal D}_{\mu}^{pq}$
 is the covariant derivative corresponding to the gauge field $A_{\mu}^p$
 (${\cal D}_{\mu}^{pq}={\delta}^{pq}{\partial}_{\mu}+f^{prq}A_{\mu}^r$).

 The gauge transformations (\ref{ftgauge}) form an abelian algebra with the
 generators $R_{\mu\nu\alpha}^{pq}$. These gauge transformations are
 not all independent, namely, for $\xi^p_\nu = {\cal D}^{qp}_\nu \xi^p$
 they vanish on-shell. Therefore, at the extremals of the action
 (\ref{fts}) the generators ${\cal R}^{pq}_{\mu\nu \alpha}$ have  zero
 modes ${\cal Z}_{\mu}^{pq}\equiv {\cal D}_{\mu}^{pq}$,
\begin{equation}
\label{ftreduce}
 {\cal R}_{1\mu\nu}^{pq}\equiv
 {\cal R}_{\mu\nu\alpha}^{pr}{\cal Z}^{rq\alpha} =
 {\varepsilon}_{\mu\nu\alpha\beta}f^{prq} \frac{\delta S_{\rm cl}}{\delta
 H_{\alpha\beta}^r}\,,
\end{equation}
 which, in their turn, are linearly independent. According to the generally
 accepted terminology, the model (\ref{fts}), (\ref{ftgauge}) and
 (\ref{ftreduce}) is an abelian gauge theory of first stage reducibility.

 Note that the gauge structure   of the FT model \cite{FT} is similar 
 to that of the Witten string \cite{Witten}.
 The FT model also has been proved to be a convenient conceptual laboratory
 for the study of the $S$-matrix unitarity in the framework of
 covariant quantization  \cite{LT}. There, it was shown that the 
 application of
 the BV quantization rules  to the model leads to a
 physically unitary theory being equivalent to a non-linear $\sigma$-model
 in
 $d=4$ dimensions \cite{FT}. Note also that various aspects of the
 quantization of the
 FT model in the framework of standard BRST symmetry have
 been discussed in Refs.~\cite{aspects,TM}.

 Now let us consider the reducible gauge model (\ref{fts}), (\ref{ftgauge})
 and (\ref{ftreduce}) in the framework of the modified triplectic 
 quantization.

 To this end, we first introduce the complete configuration space $\phi^A$,
 which is constructed according to the standard prescriptions of the
 $Sp(2)$-covariant formalism \cite{Sp2} for reducible gauge theories. Namely,
 the space of the variables $\phi^A$ consists of the initial classical fields
 $A^{p\mu}$ and $H^{p\mu\nu}$, supplemented, firstly, by $Sp(2)$-doublets of
 Faddeev--Popov ghosts, $C_{\mu}^{pa}$, introduced according to the gauge
 parameters ${\xi}_{\mu}^p$ in eq.~(\ref{ftgauge}); secondly, by additional
 sets of first-stage ghost fields, $C^{pab}$, being symmetric $Sp(2)$ tensors,
 introduced according to the gauge parameters ${\xi}^p$ for the
 generators $R_{1\mu\nu}^{pq}$ in eq.~(\ref{ftreduce}); and, finally, by sets
 of auxiliary fields (Lagrange multipliers) $B_\mu^p$, corresponding to the
 gauge parameters ${\xi}_{\mu}^p$, and first-stage $Sp(2)$-doublets $B^{pa}$,
 corresponding to the parameters ${\xi}^p$.

 The fields $\phi^A$ of the complete configuration space take values in the
 adjoint representation of a non-abelian gauge group:\footnote{
 In the following the group 
 index $p=1,\ldots,N$ will be omitted}
\[
 \phi^A=(A^{\mu},\,H^{\mu\nu};\;B^{\mu},\,B^{a};\;C^{\mu a},\,C^{ab}).
\]
 The Grassmann parities of the fields $\phi^A$ are given by
\[
 \varepsilon(A^{\mu})=\varepsilon(H^{\mu\nu})=\varepsilon(B^{\mu})=
 \varepsilon(C^{ab})=0,
\qquad
 \varepsilon(B^{a})=\varepsilon(C^{\mu a})=1.
\]

  In accordance with the quantization rules \cite{GGL}, the set of
  the fields  $\phi^A$ is supplemented by the corresponding sets of
  variables ${\phi}_{Aa}^*$, $\pi^{Aa}$ and $\bar{\phi}_A$,
\begin{eqnarray*}
 \phi_{Aa}^*&=&(A_{\mu a}^*,H_{\mu\nu a}^*;\;B_{\mu a}^*,
               B_{a|b}^*;\;C_{\mu a|b}^*,C_{a|bc}^*),\\
 \pi^{Aa}&=&(\pi_{(A)}^{\mu a},\,\pi_{(H)}^{\mu\nu a};\;\pi_{(B)}^{\mu a},
            \,\pi_{(B)}^{a|b};\;\pi_{(C)}^{\mu a|b},\,\pi_{(C)}^{a|bc}),\\
 \bar{\phi}_A&=&(\bar{A}_{\mu},\bar{H}_{\mu\nu};\;\bar{B}_{\mu},\bar{B}_{a};
              \; \bar{C}_{\mu a},\bar{C}_{ab}),
\end{eqnarray*}
 as well as by the auxiliary variables $\lambda^A$,
\[
 \lambda^A=(\lambda_{(A)}^{\mu},\,\lambda_{(H)}^{\mu\nu};\;
 \lambda_{(B)}^{\mu},\,\lambda_{(B)}^{a};\;\lambda_{(C)}^{\mu a},\,
 \lambda_{(C)}^{ab}),
\]
 with the following Grassmann parities
\begin{equation}
\label{parity}
 \varepsilon(\phi_{Aa}^*)=\varepsilon(\pi^{Aa})=\varepsilon(\phi^A)+1,
 \;\;\;
 \varepsilon(\bar{\phi}_A)=\varepsilon({\lambda}^A)=\varepsilon(\phi^A).
\end{equation}

The ghost number is assigned to the fields and auxiliary variables
by the rule that $a=1$ and $a=2$ bears ghost number +1 and
--1, respectively, for {\em upper} indices, as well as  --1 and +1,
respectively, for {\em lower} indices.
The ``external'' index $a$ on the variables $\phi^*_{Aa}$ and 
$\pi^{Aa}$ is independent from the (symmetrized) ``internal'' ones
and, therefore, separated by a vertical stroke $''|''$.

 An explicit solution of the generating equations (\ref{S}) for the
 model in question can be found in a closed form as follows:
\begin{eqnarray}
 W\!\!&=&\!\!\int{d^4}x\left(-\frac{1}{4}{\varepsilon}^{\mu\nu\rho\sigma}
     F_{\mu\nu} H_{\rho\sigma}+\frac{1}{2} A_{\mu} A^{\mu}\right)
     \nonumber\\
  &&\!\!\!+\int d^4x\,\bigg\{H_{\mu\nu a}^\ast({\cal D}^\mu C^{\nu a}-
     {\cal D}^\nu C^{\mu a})-\varepsilon^{ab}C_{\mu a|b}^\ast B^{\mu}
     +\bar{H}_{\mu\nu}({\cal D}^\mu B^{\nu}-{\cal D}^\nu B^{\mu})\nonumber\\
  &&\qquad\quad
 +C_{\mu a|b}^\ast {\cal D}^\mu C^{ab}
     -\varepsilon^{ab}C_{a|bc}^\ast B^c
     -\frac{1}{2}B_{\mu a}^\ast{\cal D}^\mu B^a+\bar{C}_{\mu a}{\cal D}^\mu
     B^a\nonumber\\
  &&\qquad\quad
 +\frac{1}{2}\varepsilon^{\mu\nu\rho\sigma}(H_{\mu\nu a}^\ast\wedge
      H_{\rho\sigma b}^\ast)C^{ab}-\frac{1}{2}\varepsilon^{\mu\nu\rho\sigma}
      (H_{\mu\nu a}^\ast\wedge\bar{H}_{\rho\sigma}) B^a\bigg\},
\label{ftS}
\end{eqnarray}
 where, after having omitted the gauge indices, we use the notations
 $A^pB^p=AB$, ${\cal D}_{\mu}B=\partial_{\mu}B+A_{\mu}\land B$,
 $(A\land B)^p=f^{pqr}A^qB^r$. It is easy to see that  
 $\Delta^a W\equiv 0$ holds.  This action, after appropriate redefinitions 
 of variables, coincides with the corresponding action obtained in \cite{BCG}
 (see eq.~(31) therein). Note also that the antifields related to $H$ appear 
 bilinear in that action. 

 A minimal admissible solution of the generating equation (\ref{X}) 
for the gauge-fixing
 functional $X$ can be represented as
\begin{eqnarray}
 X&=&\int d^4x\bigg\{\bar{A}_\mu\lambda_{(A)}^{\mu}+(\bar{H}_{\mu\nu}
    +\frac{\alpha}{2}H_{\mu\nu})\lambda_{(H)}^{\mu\nu}
    +\bar{B}_\mu\lambda_{(B)}^{\mu}+\bar{B}_a\lambda_{(B)}^{\mu}
    +\left(\bar{C}_{\mu a}
    -\beta\varepsilon_{ab}C^b_\mu\right)\lambda_{(C)}^{\mu a}\nonumber\\
    &&\qquad\quad+\bar{C}_{ab}\lambda_{(C)}^{ab}
    +\frac{\alpha}{4}\pi^a_{(H)}{}_{\mu\nu}\pi_{(H)}^{\mu\nu b}
    -\frac{\beta}{2}\varepsilon_{ab}\varepsilon_{cd}
    \pi_{(C)}{}_{\mu}^{a|c}\pi_{(C)}^{\mu b|d}\bigg\},
\label{ftX}
\end{eqnarray}
with $\alpha$ and $\beta$ being constant parameters. Here, let us remark
that the functional $F$, eq.~(\ref{F}), has been chosen as
 $F(H^{\mu\nu},C^{\mu a})=-
 \int d^4x\left(\frac{1}{4}\alpha H_{\mu\nu}H^{\mu\nu}
 +\frac{1}{2}\beta\varepsilon_{ab}C^a_\mu C^{b\mu}\right)$.

 Now, substituting the solutions of $W$, eq.~(\ref{ftS}),
 and $X$, eq.~(\ref{ftX}), into eq.~(\ref{Z}), we obtain the
 corresponding vacuum functional $Z$, with the integrand
 according to eq.~(\ref{exBRST}) being invariant under the following
 symmetry transformations:
\begin{eqnarray}
 \delta A^\alpha&=&\pi_{(A)}^{\alpha a}\mu_a,\nonumber\\
 \delta H^{\alpha\beta}&=&\pi_{(H)}^{\alpha\beta a}\mu_a
                        -\bigg({\cal D}^{[\alpha} C^{\beta] a}
                          +\varepsilon^{\alpha\beta\gamma\delta}
                          H^*_{\gamma\delta b}\wedge C^{ab}
-\frac{1}{2}\varepsilon^{\alpha\beta\gamma\delta}
                          \bar{H}_{\gamma\delta}\wedge B^a
                          \bigg)\mu_a,\nonumber\\
 \delta B^\alpha&=&\pi_{(B)}^{\alpha a}\mu_a
+\frac{1}{2}{\cal D}^\alpha B^a \mu_a,\nonumber\\
 \delta B^a&=&\pi_{(B)}^{b|a}\mu_b,\nonumber\\
 \delta C^{\alpha a}&=&\pi_{(C)}^{\alpha b|a}\mu_b
-\bigg(\varepsilon^{ab}B^{\alpha}+ {\cal D}^\alpha C^{ab}\bigg)\mu_b
                      ,\nonumber\\
 \delta C^{ab}&=&\pi_{(C)}^{c|ab}\mu_c+
 \frac{1}{2}\varepsilon^{c\{a}B^{b\}}\mu_c
                 ,\nonumber\\
		 \nonumber\\
 \delta A^*_{\alpha a}&=&\mu_a\bigg(A_{\alpha}-\frac{1}{2}
                         \varepsilon_{\alpha\beta\gamma\delta}
                         {\cal D}^\beta H^{\gamma\delta}
                         -2H^*_{\alpha\beta b}\wedge C^{\beta b}
                         -2\bar{H}_{\alpha\beta}\wedge B^\beta
                         \nonumber\\
                         &&-C^*_{\alpha b|c}\wedge C^{bc}
                         +\Big(\frac{1}{2}B^*_{\alpha b}
                         -\bar{C}_{\alpha b}\Big)\wedge B^b\bigg),\nonumber\\
                         \delta H^*_{\alpha\beta a}&=&-\mu_a\bigg(\frac{1}{4}
                         \varepsilon_{\alpha\beta\gamma\delta}
                         F^{\gamma\delta}+\frac{\alpha}{2}
                         \lambda_{(H)}{}_{\alpha\beta}\bigg),\nonumber\\
 \delta B^*_{\alpha a}&=&\mu_a\bigg(2{\cal D}^\beta\bar{H}_{\alpha\beta}
                         -\varepsilon^{bc}C^*_{\alpha b|c}\bigg),\nonumber\\
 \delta B^*_{a|b}&=&\mu_a\bigg(\varepsilon^{cd}C^*_{c|bd}
                    +\frac{1}{2}{\cal D}^\alpha B^*_{\alpha b}
                    -{\cal D}^\alpha\bar{C}_{\alpha b}
                    -\frac{1}{2}\varepsilon^{\alpha\beta\gamma\delta}
                    H^*_{\alpha\beta b}\wedge\bar{H}_{\gamma\delta}\bigg),
                    \nonumber\\
 \delta C^*_{\alpha a|b}&=&\mu_a\bigg(2{\cal D}^\beta H^*_{\alpha\beta b}
                           +\beta\varepsilon_{bc}\lambda_{(C)}^c{}_\alpha
                           \bigg),\nonumber\\
 \delta C^*_{a|bc}&=&\mu_a\bigg(-\frac{1}{2}{\cal D}^\alpha
                     {C}^*_{\alpha\{b|c\}}
                     +\frac{1}{2}\varepsilon^{\alpha\beta\gamma\delta}
                     H^*_{\alpha\beta b}\wedge H^*_{\gamma\delta |c}\bigg),
                     \nonumber\\
		     \nonumber\\
 \delta\pi_{(A)}^{\alpha a}&=&-\varepsilon^{ab}\lambda^\alpha_{(A)}\mu_b,
                              \nonumber\\
 \delta\pi_{(H)}^{\alpha\beta a}&=&
                          -\varepsilon^{ab}\lambda_{(H)}^{\alpha\beta}\mu_b
                          +\varepsilon^{ab}\bigg({\cal D}^{[\alpha}
                          B^{\beta]} 
                          +\frac{1}{2}
                          \varepsilon^{\alpha\beta\gamma\delta}
                          H^*_{\delta\gamma c}\wedge B^c
                          \bigg)\mu_b,
                          \nonumber\\
 \delta\pi_{(B)}^{\alpha a}&=&-\varepsilon^{ab}\lambda_{(B)}^\alpha\mu_b,
                              \nonumber\\
 \delta\pi_{(B)}^{a|b}&=&-\varepsilon^{ac}\lambda_{(B)}^b\mu_c,
\nonumber\\
 \delta\pi_{(C)}^{\alpha a|b}&=&
-\varepsilon^{ac}\lambda_{(C)}^{\alpha b}\mu_c
+\varepsilon^{ac}{\cal D}^\alpha B^b\mu_c
                          ,\nonumber\\
 \delta\pi_{(C)}^{a|bc}&=&-\varepsilon^{ad}\lambda_{(C)}^{bc}\mu_d,
                          \nonumber\\
			  \nonumber\\
 \delta\bar{A}_\alpha&=&\mu_a\varepsilon^{ab}A^*_{\alpha b},\nonumber\\
 \delta\bar{H}_{\alpha\beta}&=&\mu_a\bigg(\varepsilon^{ab}H^*_{\alpha\beta b}
                               -\frac{\alpha}{2}\pi_{(H)}^a{}_{\alpha\beta}
                               \bigg),\nonumber\\
 \delta\bar{B}_\alpha&=&\mu_a\varepsilon^{ab}B^*_{\alpha b},\nonumber\\
 \delta\bar{B}_a&=&\mu_c\varepsilon^{cb}B^*_{b|a},\nonumber\\
 \delta\bar{C}_{\alpha a}&=&\mu_c\bigg(\varepsilon^{cb}C^*_{\alpha b|a}
                            +\beta\varepsilon_{ab}\pi_{(C)}^{c|b}{}_\alpha
                            \bigg),\nonumber\\
\label{ftBRST1}
 \delta\bar{C}_{ab}&=&\mu_c\varepsilon^{cd}C^*_{d|ab},
\end{eqnarray}
 where symmetrization and antisymmetrization is taken as 
 $A^{\{ab\}}=A^{ab}+A^{ba}$ and $A^{[\alpha\beta]}=
 A^{\alpha\beta}-A^{\beta\alpha}$, respectively.
 Eqs.~(\ref{ftBRST1}) realize the transformations of extended BRST symmetry
 of the vacuum functional in terms of the anticanonically conjugated variables
 $(\phi^A$, $\phi^*_{Aa})$ and $(\pi^{Aa}$, $\bar{\phi}_A)$.

 Integrating in eq.~(\ref{Z}) over the variables $\phi^*_{Aa}$,
 $\pi^{Aa}$, $\bar{\phi}_A$ and $\lambda^A$, we represent the
 vacuum functional $Z$ as an integral over the fields $\phi^A$
 of the complete configuration space,
\begin{equation}
\label{ftZ}
 Z=\int d\phi\,\Delta\exp\bigg\{\frac{i}{\hbar}S_{\rm eff}^{(0)}(\phi)
 \bigg\},
\end{equation}
 where
\begin{eqnarray}
 S_{\rm eff}^{(0)}&=&\int{d^4}x\left(-\frac{1}{4}
                     \varepsilon^{\mu\nu\rho\sigma}F_{\mu\nu}
                     H_{\rho\sigma}+\frac{1}{2} A_{\mu} A^{\mu}\right)
                     \nonumber\\
 &&+\int\,d^4x\left\{\frac{\alpha}{4}G_{\mu\nu}^a{\cal M}^{-1}_{ab}
                     {\cal K}_c^{b[\mu\nu][\rho\sigma]}G_{\rho\sigma}^c
                     -\frac{\beta}{2}\varepsilon_{ab}
                     \varepsilon_{cd}
\Big({\cal D}_\mu C^{ac}\Big)\Big({\cal D}^\mu C^{bd}\Big)
                     \right\}\nonumber\\
\label{ftS2}
                  &&+\int\,d^4x\left(\alpha B_\mu {\cal D}_\nu H^{\nu\mu}+
                     \beta(\varepsilon_{ab} B^a {\cal D}_\mu C^{\mu b}
                     - B_\mu B^\mu)\right),\\
\nonumber\\
\label{ftDelta}
 \Delta&=&\int dH^\ast\exp\left\{\frac{2i}{\alpha\hbar}\int
          d^4x\,H_{0ia}^\ast {\cal M}^{ab} H_{0jb}^\ast \eta^{ij}\right\}.
\end{eqnarray}

 In eq.~(\ref{ftS2}), we have used the following notations:
\begin{eqnarray}
 {\cal K}_b^{a[\mu\nu][\rho\sigma]} &\equiv& \frac{1}{2}\{\delta_b^a(
 \eta^{\mu\rho}\eta^{\nu\sigma}-\eta^{\mu\sigma}\eta^{\nu\rho})+
 \alpha {\cal C}_b^a\varepsilon^{\mu\nu\rho\sigma}\},
\nonumber\\
\label{KG}
 G_{\mu\nu}^a &\equiv& \left({\cal D}_\mu C_\nu^a-{\cal D}_\nu C_\mu^a\right)
 -\frac{\alpha}{4}\varepsilon_{\mu\nu\rho\sigma} {\cal B}^a H^{\rho\sigma}.
\end{eqnarray}
 The matrix ${\cal M}^{-1}_{ab}$ in (\ref{ftDelta})
is the inverse of ${\cal M}^{ab}$,
\begin{eqnarray}
\label{M^ab}
 {\cal M}^{ab} \equiv \varepsilon^{ab}
 -\alpha^2{\cal C}_c^a{\cal C}_d^b\varepsilon^{cd},
 \qquad
 {\cal M}^{ac}{\cal M}^{-1}_{cb}=\delta_b^a;
\end{eqnarray}
 here the matrices ${\cal C}_b^a$ and ${\cal B}^a$ are defined by
\begin{eqnarray}
\label{CB} 
{\cal C}_b^a E \equiv \varepsilon_{bc} C^{ac}\wedge E ,
\qquad
  {\cal B}^a E\equiv B^a\wedge E.
\end{eqnarray}

 The functional $S_{\rm eff}^{(0)}$ in eq.~(\ref{ftS2}) is the tree
 approximation to the gauge-fixed quantum action of the theory, while the
 functional $\Delta$ in eq.~(\ref{ftDelta}) can be considered as a
 contribution to the integration measure. (The proof of the above
 representation of the vacuum functional (\ref{ftZ}) is given in the 
 Appendix).

 The integrand in eq.~(\ref{ftZ}) is invariant under the following symmetry
 transformations:
\begin{eqnarray}
 \delta A_{\mu}&=&0,\nonumber\\
 \delta H^{\alpha\beta}
&=&-\varepsilon^{ab}{\cal M}^{-1}_{bc}{\cal K}_d^{c[\alpha\beta]
 [\gamma\delta]}G_{\gamma\delta}^d\mu_a,\nonumber\\
 \delta B^\alpha&=&\frac{1}{2} {\cal D}^\alpha B^a\mu_a,\nonumber\\
\delta B^a&=&0, \nonumber\\
 \delta C^{\alpha a}&=&({\cal D}^\alpha C^{ab}-\varepsilon^{ab}B^
 \alpha)\mu_b,\nonumber\\
\label{ftBRST2}
 \delta C^{ab}&=&\frac{1}{2}B^{\{a}\varepsilon^{b\}c}\mu_c.
\end{eqnarray}
 These transformations are the (anti)BRST transformations 
 of the $Sp(2)$ invariant
 action $S_{\rm eff}^{(0)}$ which, together with the integration 
 measure $d\phi\Delta$ in eq.~(\ref{ftZ}), is left invariant:
\[
 \delta(d\phi)=d\phi\,\delta^4(0)\int d^4x\,{\rm Tr}\,{\cal W},
\]
\[
 \delta\Delta=-\Delta\delta^4(0)\int d^4x\,{\rm Tr}\,{\cal W},
\]
\[
 \delta\left(\exp\left\{\frac{i}{\hbar}S_{\rm eff}^{(0)}\right\}\right)=0,
\]
 where the following notations have been used:
\[
  {\cal W}=-3\alpha^2\varepsilon^{ab}
  ({\cal M}^{-1}_{bc}{\cal C}_d^c{\cal B}^d)\mu_a\,,
  \,\,\,\,{\rm Tr}\,{\cal W}\equiv\sum_{p=1}^N {\cal W}^{pp}.
\]

 Consequently, eqs.~(\ref{ftBRST2}) realize the transformations of extended 
 BRST symmetry for the vacuum functional (\ref{ftZ})
 in terms of the variables $\phi^A$ of the complete configuration space. 
 Remarkably, the (anti)BRST transformations of the 
 classical field $H^{\alpha\beta}$ essentially depend on the gauge 
 parameter $\alpha$ whereas all the others coincide -- up to the $\pi$-terms 
 -- with the transformations (\ref{ftBRST1}).\footnote{
To observe this fact, it is sufficient to make the replacements 
$\mu_a \rightarrow -\mu_a$ and $B^\mu \rightarrow - B^\mu$.
 } 
 The appearance of the parameter $\alpha$ can be traced back to the
 non-linear dependence of the extended action $S$ on the antifields 
 $H^*_{\mu\nu a}$.

 Note that, taking into account the action (\ref{ftS2}) and the contribution 
 to the integration measure (\ref{ftDelta}),  
 the vacuum functional (\ref{ftZ}) obtained for the Freedman--Townsend model 
 leads to
 the unitarity \cite{LM} of the physical $S$ matrix (for discussions of the
 unitarity problem in the case of this model, see also \cite{aspects,CLL,LT}).

 For the first time, the covariant quantization of the Freedman--Townsend 
 model
 in the framework of extended BRST invariance has been performed by Barnich,
 Constantinescu and Gregoire \cite{BCG}.
 However, these authors  used a more complicated, $Sp(2)-$noncovariant 
 method of gauge fixing which might have prevented them from explicitly 
 solving their expression with respect to the fields $\phi^A$ completely. 
 Namely, they left the dependence
 of the action from the antifields $H^*_{\alpha\beta}$ and
 ${\bar H}_{\alpha\beta}$ which we were able to integrate out. 
 Similar results have been obtained in Ref. \cite{ACA}, where it has been
 shown that the elimination of antifields leads to a nonlocal action in terms
 of the fields. The same conclusion could be drawn from our result since the
 matrix ${\cal M}^{-1}$ is given as an infinite series only.

 \section{$\mbox{$W_2$}-$gravity}
 \setcounter{equation}{0}

 The model of $W_2-$gravity \cite{W2} is described by the action
\begin{equation}
 \label{w2s}
 S_{cl}(\varphi,h)=\frac{1}{2\pi}\int{d^2}z \left(\partial\varphi
 \bar{\partial}\varphi-h(\partial\varphi)^2\right),
\end{equation}
 where $\varphi$ and $h$ are bosonic classical fields,
 $\varepsilon(\varphi)=\varepsilon(h)=0$,
 defined on a space with complex coordinates, $(z,\bar{z})$, so that
 $\partial={\partial}/{\partial z}\,$,
$\bar{\partial}={\partial}/{\partial\bar{z}}\,$.

 The  action (\ref{w2s}) is invariant under the gauge transformations
\begin{equation}
 \label{w2gauge}
\begin{array}{rcl}
 \delta\varphi&=&(\partial\varphi)\xi,\\
 \\
 \delta h&=&\bar{\partial}\xi-h\partial\xi+(\partial h)\xi
\end{array}
\end{equation}
 with the gauge function $\xi(z, {\bar z})$.
 These transformations form a closed algebra,
\begin{equation}
 \label{w2alg}\\
\begin{array}{cl}
 &{[}\delta_{\xi(1)},\;\delta_{\xi(2)}{]}=\delta_{\xi(1,2)}\,,\\
 \\
 &\xi_{(1,2)}=(\partial\xi_{(1)})\xi_{(2)}-(\partial\xi_{(2)})\xi_{(1)}.
 \nonumber
\end{array}
\end{equation}

 Note that the quantum properties of $W_2-$gravity, considered within the BV
 method \cite{BV}, have been discussed in \cite{JST,AB,DeJ}, where also the 
 one-loop  anomaly has been calculated. Recently, the quantization of that 
 model has been
 performed in the triplectic formalism \cite{BG}, however, with another kind
of gauge fixing being different from using the gauge fixing functional $X$.

 Now, we consider the gauge model (\ref{w2s}), (\ref{w2gauge}) and
 (\ref{w2alg}) in the framework of the modified triplectic quantization.
 First,
 let us introduce the complete configuration space $\phi^A$, whose structure
 in the case of the model in question is determined by the rules of the
 $Sp(2)$ formalism  for irreducible gauge theories. Thus, the space
 of the variables $\phi^A$ is constructed by supplementing the initial space
 of the fields $(\varphi,h)$ with the doublet $C^a$, $\varepsilon(C^a)=1$, of
 Faddeev--Popov ghosts, and the Lagrange multiplier $B$, $\varepsilon(B)=0$,
 corresponding to the gauge parameter $\xi$ in eq.~(\ref{w2gauge}).

 The fields $\phi^A$ of the complete configuration space,
\[
 \phi^A=(\varphi,\;h;\;B,\;C^a),
\]
 are supplemented by the sets of the variables $\phi^*_{Aa}$, $\pi^{Aa}$
 and $\bar{\phi}_A$,
\begin{eqnarray*}
 \phi^*_{Aa}&=&(\varphi^*_a,\;h^*_a;\;B^*_a,\;C^*_{a|b}),\\
 \pi^{Aa}&=&(\pi_{(\varphi)}^a,\;\pi_{(h)}^a;\;\pi_{(B)}^a,\;
 \pi_{(C)}^{a|b}),\\
 \bar{\phi}_A&=&(\bar{\varphi},\;\bar{h};\;\bar{B},\;\bar{C}_a),
\end{eqnarray*}
 as well as by the additional variables $\lambda^A$,
\[
 \lambda^{A}=(\lambda_{(\varphi)},\;\lambda_{(h)};\;\lambda_{(B)},\;
 \lambda_{(C)}^{a}),
\]
 with the Grassmann parities given by eq.~(\ref{parity}).

 An action functional $S$ of the gauge model (\ref{w2s}), (\ref{w2gauge})
 and (\ref{w2alg}) satisfying the generating equations (\ref{S}) in tree 
 approximation, i.e.,~when ignoring their r.h.s., is given as follows:
\begin{eqnarray}
 S\!\!&=&\!\!\frac{1}{2\pi}\int{d^2}z\left(\partial\varphi\;
     \bar{\partial}\varphi-h(\partial\varphi)^2\right)\nonumber\\
 &&\!\!\!+\int{d^2}z\bigg\{
     \varphi^*_aC^a\partial\varphi+h^*_a\left(\bar{\partial}C^a
     -h\partial C^a+C^a\partial h\right)
\nonumber\\
  &&\!\!\!+\bigg(\frac{1}{2}B^*_a-\bar{C}_a\bigg)\bigg[
     \left(C^a\partial B-B\partial C^a\right)
\label{w2S}
  +\frac{1}{6}\varepsilon_{bd}
     \bigg(C^{\{a}\big(\partial^2C^{d\}}\big)C^b
     -C^{\{a}\big(\partial C^{d\}}\big)\partial C^b\bigg)\bigg]
     \nonumber\\
  &&-C^*_{a|b}\bigg(\varepsilon^{ab}B
     +\frac{1}{2}C^{\{a}\partial C^{b\}}\bigg)
  +\bar{\varphi}\bigg(B\partial\varphi+\frac{1}{2}
     \varepsilon_{ab}C^a\partial(C^b\partial\varphi)\bigg)
  +\bar{h}\bigg(\bar{\partial}B-h\partial B+B\partial h\bigg)
\nonumber\\
  &&+\frac{1}{2}\varepsilon_{ab}\bar{h}\bigg(
     C^a\partial\left(\bar{\partial}C^b
     -h\partial C^b+C^b\partial h\right)
  +\left(\bar{\partial}C^b-h\partial C^b+C^b\partial h\right)
 \partial C^a\bigg) \bigg\}.
\end{eqnarray}
Obviously, the application of the differential operator $\Delta^a$ leads to
terms being proportional to $\delta(0)$. However, since the model is
strictly two-dimensional the dimensional regularization is not
applicable. Therefore, it is not surprising that the model has an anomaly
which can not be compensated by appropriate counterterms.

 Furthermore, a solution of the generating equations determining 
 the gauge-fixing  functional $X$ can be represented as
\begin{eqnarray}
 X&=&\int{d^2}z\bigg\{(\bar{\varphi}-\alpha\varphi-\beta h)\lambda_{(\varphi)}
     +(\bar{h}-\beta\varphi-\gamma h)\lambda_{(h)}
     +\bar{B}\lambda_{(B)}+\bar{C}_a\lambda_{(C)}^a\nonumber\\
\label{w2X}
  &&\qquad\quad - \frac{\alpha}{2}\varepsilon_{ab}
     \pi^a_{(\varphi)}\pi^b_{(\varphi)}
     -\beta\varepsilon_{ab}\pi^a_{(\varphi)}\pi^b_{(h)}
     -\frac{\gamma}{2}\varepsilon_{ab}\pi^a_{(h)}\pi^b_{(h)}\bigg\},
\end{eqnarray}
 with $\alpha$, $\beta$ and $\gamma$ being constant parameters.
 Here, $F$ has been chosen as
 $F(\varphi,h)=\int d^2z\left(\frac{1}{2}\alpha \varphi^2 + \beta \varphi h
 +\frac{1}{2}\gamma h^2\right)$.

 The vacuum functional (\ref{Z}) corresponding to 
 the solutions
 (\ref{w2S}) and (\ref{w2X}) of the generating equations that determine the
 action $W$ in tree approximation and the gauge-fixing functional $X$, is 
 invariant under
 the following transformations of extended BRST symmetry, expressed 
(for simplicity) in terms of the derivatives of $S$:
\begin{eqnarray}
 \delta\phi^A &=&\bigg(\pi^{Aa} - \frac{\delta
 S}{\delta\phi^*_{Aa}}\bigg)\mu_a, \nonumber\\
 \nonumber\\
 \delta\varphi^*_a &=&\mu_a\bigg(\frac{\delta S}{\delta\varphi}+
 \alpha\lambda_{(\varphi)}+\beta\lambda_{(h)}\bigg),\nonumber\\
 \delta h^*_a &=& \mu_a\bigg(\frac{\delta S}{\delta h}+
 \beta\lambda_{(\varphi)} + \gamma\lambda_{(h)}\bigg),\nonumber\\
 \delta B^*_a &=& \mu_a \frac{\delta S}{\delta B}, \nonumber\\
 \delta C^*_{a|b} &=& \mu_a\frac{\delta S}{\delta C^b},\nonumber\\
 \nonumber\\
 \delta\pi^{Aa}&=&\varepsilon^{ab}\bigg(\frac{\delta S}
                  {\delta\bar{\phi}^A}-\lambda^A\bigg)\mu_b,\nonumber\\
		  \nonumber\\
 \delta\bar{\varphi} &=& \mu_a(\varepsilon^{ab}\varphi^*_b +
 \alpha\pi^a_{(\varphi)}+
 \beta\pi^a_{(h)}),\nonumber\\
 \delta\bar{h} &=& \mu_a(\varepsilon^{ab}h^*_b + \beta\pi^a_{(\varphi)}
  + \gamma \pi^a_{(h)}),\nonumber\\
 \delta\bar{B} &=& \mu_a\varepsilon^{ab}B^*_b,\nonumber\\
 \delta\bar{C_a}&=&\mu_b\varepsilon^{bd}C^*_{d|a}.
 \end{eqnarray}

 Substituting the solutions (\ref{w2S}), (\ref{w2X}) for the
 action $S$ and the gauge-fixing functional $X$ into eq.~(\ref{Z}), and
 integrating out the variables $\phi^*_{Aa}$, $\pi^{Aa}$, $\bar{\phi}_A$,
 $\lambda^A$, we obtain the vacuum functional $Z$ as an integral over the
 fields $\phi^A$ of the complete configuration space,
\begin{eqnarray}
 \label{w2Z}
 Z=\int d\phi\,\exp\left\{\frac{i}{\hbar}S_{\rm eff}(\phi)\right\},
\end{eqnarray}
 where $S_{\rm eff}$ is the gauge-fixed tree approximation of the 
 quantum action
\begin{eqnarray}
 S_{\rm eff}\!\!&=\!\!&\frac{1}{2\pi}\int{d^2}z\left(\partial\varphi
     \bar{\partial}\varphi-h(\partial\varphi)^2\right)\nonumber\\
  &&+\int{d^2}z\bigg[(\alpha\varphi+\beta h)B\partial\varphi
     +(\beta\varphi+\gamma h)\left(\bar{\partial}B-h\partial B
     +B\partial h\right)\!\bigg]
      \nonumber\\
  &&+\frac{1}{2}\varepsilon_{ab}\int{d^2}z\bigg[\bigg(
     \alpha C^b\partial\varphi+\beta\left(\bar{\partial}C^b-h\partial C^b
     +C^b\partial h\right)\bigg)C^a\partial\varphi
  -(\alpha\varphi+\beta h)C^a\partial(C^b\partial\varphi)
     \nonumber\\
  &&+\bigg(\beta C^b\partial\varphi+\gamma\left(\bar{\partial}C^b
     -h\partial C^b+C^b\partial h\right)\bigg)
     \left(\bar{\partial} C^a-h\partial C^a
     +C^a\partial h\right)\nonumber\\
  &&-(\beta\varphi+\gamma h)\bigg(
     C^a\partial\left(\bar{\partial}C^b-h\partial C^b
     +C^b\partial h\right)
\label{w2S2}
     +\left(\bar{\partial}C^b-h\partial C^b
     +C^b\partial h\right)\partial C^a
     \bigg)\bigg].
\end{eqnarray}
 The quantum  action $S_{\rm eff}$, eq.~(\ref{w2S2}), and the integration
measure are invariant under
 the following (anti)BRST transformations:
\begin{eqnarray}
 \delta\varphi&=&C^a\partial\varphi\;\mu_a,\nonumber\\
 \delta h&=&\left(\bar{\partial}C^a-h\partial C^a+C^a\partial h\right)
           \mu_a,\nonumber\\
 \delta B&=&\frac{1}{2}\left(C^a\partial B-B\partial C^a\right)\mu_a
            +\frac{1}{12}\varepsilon_{bd}
            \bigg(C^{\{a}(\partial^2C^{d\}})C^b
            -C^{\{a}(\partial C^{d\}})\partial C^b\bigg)\mu_a,
            \nonumber\\
\label{w2BRST2}
 \delta C^a&=&\bigg(\varepsilon^{ab}B-\frac{1}{2}C^{\{a}\partial C^{b\}}
              \bigg)\mu_b.
\end{eqnarray}

 Thus we conclude that eqs.~(\ref{w2BRST2}) realize the transformations of
 extended BRST symmetry in terms of the variables of the complete 
 configuration space. Furthermore, introducing the action of the 
 (anti)BRST operators  $s^a$ onto the fields $\phi^A$ according to the 
 rule $\delta \phi^A = (s^a\phi^A)\mu_a$, one can rewrite the
 effective action (\ref{w2S2}) in the following compact form:
\begin{eqnarray}
 \label{w2seff}
 S_{\rm eff} = S_{\rm cl} +
\frac{1}{2}\varepsilon_{ab}s^b s^a F(\varphi, h),
\end{eqnarray}
 i.e., one obtains the usual effective action of the $Sp(2)$-covariant
 approach. The invariance of $S_{\rm eff}$ under the transformations
 (\ref{w2BRST2}) follows by virtue of the (generalized) nilpotency,
 $s^{\{a}s^{b\}} =0$.

\section{Two-dimensional Gravity with Dynamical Torsion}
\setcounter{equation}{0}

 The theory  of two-dimensional gravity with dynamical torsion is
 described in terms of the zweibein and Lorentz connection
 $(e^i_\mu,\omega_\mu^{ij} = \varepsilon^{ij}\omega_\mu)$
 by the action \cite{VK}
\begin{equation}
\label{2grs}
 S_{cl}(e^i_\mu,\omega_\mu) = \int {d^2}x\;e\bigg(\frac{1}{16\alpha}
 R_{\mu\nu}{}^{ij}R^{\mu\nu}{}_{ij}-\frac{1}{8\beta}T_{\mu\nu}{}^{i}
 T^{\mu\nu}{}_{i}-\gamma\bigg),
\end{equation}
 where $\alpha$, $\beta$ and $\gamma$ are constant parameters. In
 eq.~(\ref{2grs}), the Latin indices are lowered with the help of the
 Minkowski metric $\eta_{ij}={\rm diag}\,(+1,-1)$, and the Greek indices,
 with the help of the metric tensor $g_{\mu\nu}=\eta_{ij}e^i_\mu e^j_\nu$.
 Besides, the following notations are used:
\begin{eqnarray*}
 e&=&{\rm det}\,e^i_\mu\,,\\
 R_{\mu\nu}{}^{ij}&=&\varepsilon^{ij}\partial_\mu\omega_\nu
                   -(\mu\leftrightarrow\nu),\\
 T_{\mu\nu}{}^{i}&=&\partial_\mu e^i_\nu
                    +\varepsilon^{ij}\omega_\mu e_{\nu j}-
                    (\mu\leftrightarrow\nu),
\end{eqnarray*}
 where $\epsilon^{ij}$ is a constant antisymmetric tensor, $\epsilon^{01}=-1$.

 Note that the model (\ref{2grs}) is the most general theory of
 two-dimensional $R^2-$gravity with independent dynamical torsion that leads
 to second-order equations of motion for the zweibein and Lorentz connection.
 Thus, supplementing the action eq.~(\ref{2grs}) by the Einstein--Hilbert term
 $eR$ would not affect the classical field equations, since in two dimensions
 it reduces to a trivial total divergence.
 
 Originally, the action (\ref{2grs}) was proposed \cite{VKstring}
 in the context of bosonic string theory, where it was used to describe the 
 dynamics
 of string geometry. There, moreover, it was proved that the string with
 dynamical geometry has no critical dimension. 

 An attractive feature of the model
 (\ref{2grs}) is its complete integrability. The corresponding equations 
 of motion
 have been studied in conformal \cite{VK,K} and light-cone \cite{KSch}
 gauges.\footnote{A convenient general framework for the study of 
 this and other
 two-dimensional theories is also provided by the Poisson-$\sigma$ model 
 approach \cite{sigma-mod}.}
 It was established that the model contains solutions with constant curvature 
 and
 zero torsion, thus incorporating several other two-dimensional gravity models
 \cite{models} whose actions, however, do not have a purely geometric 
 interpretation.

 The action (\ref{2grs}) is invariant under local Lorentz rotations of the
 zweibein  $e^i_\mu$, which infinitesimally imples the gauge transformations
\begin{eqnarray}
\label{2grgauge1}
 \delta_\zeta e^i_\mu=\varepsilon^{ij}e_{\mu j}\zeta,
\qquad
 \delta_\zeta \omega_\mu=-\partial_\mu\zeta,
\end{eqnarray}
 with the parameter $\zeta$. Similarly, the general coordinate invariance
 of eq.~(\ref{2grs}) leads to the gauge transformations
\begin{eqnarray}
\label{2grgauge2}
 \delta_\xi e^i_\mu=e^i_\nu\partial_\mu\xi^\nu+(\partial_\nu
 e^i_\mu)\xi^\nu,
\qquad
 \delta_\xi\omega_\mu=\omega_\nu\partial_\mu\xi^\nu+(\partial_\nu
 \omega_\mu)\xi^\nu
\end{eqnarray}
 with the parameters $\xi^\mu$. The gauge transformations (\ref{2grgauge1})
 and (\ref{2grgauge2}) form a closed algebra
\begin{eqnarray}
\label{2gralg}
 {[}\delta_{\zeta(1)},\;\delta_{\zeta(2)}{]}&=&0\,,\nonumber\\
 {[}\delta_{\xi(1)},\;\delta_{\xi(2)}{]}&=&\delta_{\xi(1,2)}\,,\\
 {[}\delta_\zeta,\;\delta_\xi{]}&=&\delta_{\zeta^{'}}\,,\nonumber
\end{eqnarray}
 where
\[
 \xi^\mu{}_{(1,2)}=(\partial_\nu\xi^\mu{}_{(1)})\xi^\nu{}_{(2)}-
 (\partial_\nu\xi^\mu{}_{(2)})\xi^\nu{}_{(1)}\,,\;\;\;
 \zeta^{'}=(\partial_\mu\zeta)\xi^\mu.\nonumber
\]

 Note that in Ref.~\cite{II} a gauge model classically equivalent to
 (\ref{2grs}), (\ref{2grgauge1}), (\ref{2grgauge2}) and (\ref{2gralg})
 was proposed by means of artificially adding the Einstein--Hilbert term
 coupled to an additional  scalar field, $\sigma e R$; 
 however, in this equivalent formulation the algebra of the corresponding
 gauge transformations closes only on-shell.

 The Hamiltonian structure of the gauge symmetries of the original model
 was studied in Ref.~\cite{Strobl}, and its canonical quantization,
 in Ref.~\cite{quant}. Quantum properties of the model in the light-cone
 gauge were discussed in Ref.~\cite{quant_also}, proving also, despite of the
 nonpolynomial structure of the theory, its renormalizability.
 Its quantization within the $Sp(2)$ covariant approach has been
 considered in Refs.~\cite{PP}.

 Now, we consider the gauge model (\ref{2grs}), (\ref{2grgauge1}), 
 (\ref{2grgauge2})
 and (\ref{2gralg}) in the framework of the modified triplectic formalism
 \cite{GGL}.

 The complete configuration space $\phi^A$, constructed by the rules
 of the $Sp(2)$ covariant quantization of irreducible theories, consists
 of the initial classical fields $(e^i_\mu,\omega_\mu)$, the doublets of
 the Faddeev--Popov ghosts ($C^a$, $C^{\mu a}$) and the Lagrangian
 multipliers ($B$, $B^\mu$) introduced according to the number of the
 gauge parameters in eqs.~(\ref{2grgauge1}) and (\ref{2grgauge2}), i.e.,
 $\zeta$ and $\xi^\mu$, respectively. The Grassmann parities
 of the fields $\phi^A$,
\[
 \phi^A=(e^i_\mu,\;\omega_\mu;\;B,\;B^\mu;\;C^a,\;C^{\mu a}),
\]
 are given by
\[
 \varepsilon(e^i_\mu)=\varepsilon(\omega_\mu)=\varepsilon(B)=
 \varepsilon(B^\mu)=0,
\qquad
 \varepsilon(C^a)=\varepsilon(C^{\mu a})=1.
\]

 The fields $\phi^A$ of the complete configuration space are supplemented 
 by the
 following sets of the variables $\phi^*_{Aa}$, $\pi^{Aa}$, $\bar{\phi}_A$
 and $\lambda^{A}$:
\begin{eqnarray*}
 \phi^*_{Aa}&=&(e^{*\mu}_{ia},\;\omega^{*\mu}_a;\;B^*_a,\;B^*_{\mu a};\;
 C^*_{a|b},\;C^*_{\mu a|b}),\\
 \pi^{Aa}&=&(\pi_{(e)}^{ia}{}_\mu,\;\pi_{(\omega)}^a{}_\mu;\;\pi_{(B)}^a,\;
            \pi_{(B)}^{\mu a};\;\pi_{(C)}^{a|b},\;\pi_{(C)}^{\mu a|b}),\\
 \bar{\phi}_A&=&(\bar{e}^\mu_i,\;\bar{\omega}^\mu;\;\bar{B},\;
                \bar{B}_\mu;\;\bar{C}_a,\;\bar{C}_{\mu a}),\\
 \lambda^A&=&(\lambda_{(e)}^i{}_\mu,\;\lambda_{(\omega)}{}_\mu;\;
             \lambda_{(B)},\;\lambda_{(B)}^\mu;\;\lambda_{(C)}^a,\;
             \lambda_{(C)}^{\mu a}).
\end{eqnarray*}

 Again, we are faced with the problem that the model is stricly
two-dimensional and, therefore, the regularization by setting $\delta(0)=0$  
 will not be applicable. A functional that satisfies the generating
equations (\ref{S}) for the classical action $S$ in this case of the gauge
model (\ref{2grs}), (\ref{2grgauge1}),
 (\ref{2grgauge2}) and (\ref{2gralg}) can be found in a closed form as
follows:
\begin{eqnarray}
 \!\!\!\!\!&\!\!S\!\!&=
\int {d^2}x\;e\bigg(\frac{1}{16\alpha}
     R_{\mu\nu}{}^{ij}R^{\mu\nu}{}_{ij}-\frac{1}{8\beta}T_{\mu\nu}{}^{i}
     T^{\mu\nu}{}_{i}-\gamma\bigg)
\nonumber\\
  &&\!\!\!
  +\int d^2x\bigg\{e^{*\mu}_{ia}\left(
     \varepsilon^{ij}e_{\mu j}C^a+
     C^{\lambda a}\partial_\lambda e^i_\mu
     +e^i_{\lambda}\partial_\mu C^{\lambda a}\right)
  +\omega^{*\mu}_a\left(-\partial_\mu C^a+C^{\lambda a}
     \partial_\lambda\omega_\mu+\omega_\lambda\partial_\mu C^{\lambda a}
     \right)\nonumber\\
  &&\!\!\!
  +\left(\hbox{\large$\frac{1}{2}$}
  	 B^*_a-\bar{C}_a\right)\!\!\bigg[C^{\mu a}\partial_\mu B
     	-B^\mu\partial_\mu C^a 
  +\frac{1}{6}\varepsilon_{bd}\left(
     C^{\lambda b}\partial_\lambda(C^{\mu\{a}\partial_\mu C^{d\}})
     -(C^{\mu\{a}\partial_\mu C^{\lambda d\}})\partial_\lambda C^{b}\right)
     \bigg]\nonumber\\
  &&\!\!\!
  +\!\left(\hbox{\large$\frac{1}{2}$} 	
  	B^*_{\mu a}\!-\!\bar{C}_{\mu a}\!\right)\!\!
     	\bigg[C^{\lambda a}\partial_\lambda B^\mu
     	\!-\!B^\lambda\partial_\lambda C^{\mu a} 
  \!+\!\hbox{\large$\frac{1}{6}$}
  \varepsilon_{bd}
     \left[C^{\sigma b}\partial_\sigma(C^{\lambda\{a}
     \partial_\lambda C^{\mu d\}})
     \!-\!(C^{\lambda\{a}\partial_\lambda
     C^{\sigma d\}})\partial_\sigma C^{\mu b}\right]\bigg]\nonumber\\
  &&\!\!\!
  -C^*_{a|b}\left(\varepsilon^{ab}B
     +\hbox{\large$\frac{1}{2}$}
     C^{\mu \{a}\partial_\mu C^{b\}}\right)
     -C^*_{\mu a|b}\left(\varepsilon^{ab}B^\mu
     +\hbox{\large$\frac{1}{2}$}
     C^{\lambda \{a}\partial_\lambda C^{\mu b\}}\right)
     \nonumber\\
  &&\!\!\!
  +\bar{e}^\mu_i\bigg[\varepsilon^{ij}Be_{\mu j}
     +B^\lambda\partial_\lambda e^i_\mu
     +e^i_\lambda\partial_\mu B^\lambda
  +\hbox{\large$\frac{1}{2}$}
  \varepsilon_{ab}\bigg(\!\!
  \left(e^i_\mu C^b+\varepsilon^{ij}
     C^{\lambda b}\partial_\lambda e_{\mu j}
     +\varepsilon^{ij}e_{\lambda j}\partial_\mu
     C^{\lambda b}\right)C^a\nonumber\\
  &&\!\!\!
  -C^{\lambda a}\partial_\lambda\left(\varepsilon^{ij}e_{\mu j}C^b
     +C^{\sigma b}\partial_\sigma e^i_\mu+e^i_\sigma\partial_\mu
      C^{\sigma b}\right)
  +\left(\varepsilon^{ij}e_{\lambda j}C^b
     +(\partial_\sigma e^i_\lambda)C^{\sigma b}
     +e^i_\sigma\partial_\lambda C^{\sigma b}\right)\partial_\mu C^{\lambda a}
     \bigg)\bigg]\nonumber\\
  &&\!\!\!
  +\bar{\omega}^\mu\bigg[
     -\partial_\mu B+B^\lambda\partial_\lambda\omega_\mu
     +\omega_\lambda\partial_\mu B^\lambda
  -\hbox{\large$\frac{1}{2}$}
  \varepsilon_{ab}\bigg(
     C^{\lambda a}\partial_\lambda
     \left(C^{\sigma b}\partial_\sigma\omega_\mu+\omega_\sigma\partial_\mu
     C^{\sigma b}-\partial_\mu C^b\right)\nonumber\\
\label{2grS}
  &&\!\!\!
  -\left(C^{\sigma b}\partial_\sigma\omega_\lambda
     +\omega_\sigma\partial_\lambda C^{\sigma b}-\partial_\lambda C^b\right)
 \partial_\mu C^{\lambda a}\bigg)\bigg]\bigg\}.
\end{eqnarray}

 A solution of the generating equations determining the gauge-fixing
 functional $X$ can be chosen as
\begin{eqnarray}
 X\!\!&=&\!\!\int d^2x\bigg\{\bigg(\bar{e}^\mu_i-
 p\eta^{\mu\nu}\eta_{ij}e^j_\nu
     \bigg)\lambda_{(e)}^i{}_{\mu}
     +\bigg(\bar{\omega}^\mu-q\eta^{\mu\nu}\omega_\nu\bigg)
     \lambda_{(\omega)}{}_{\mu}
     +\bar{B}\lambda_{(B)}
     +\bar{B}_{\mu}\lambda_{(B)}^{\mu}\nonumber\\
\label{2grX}
   &&\quad\quad+\bar{C}_a\lambda_{(C)}^a
     +\bar{C}_{\mu a}\lambda_{(C)}^{\mu a}
     -\frac{p}{2}\varepsilon_{ab}\eta_{ij}\eta^{\mu\nu}
     \pi^{ia}_{(e)}{}_{\mu}\pi^{jb}_{(e)}{}_{\nu}
     -\frac{q}{2}\varepsilon_{ab}\eta^{\mu\nu}
     \pi^{a}_{(\omega)}{}_{\mu}\pi^{b}_{(\omega)}{}_{\nu}\bigg\},
\end{eqnarray}
 where
 $F(e^i_\mu, \omega_\mu)=\frac{1}{2} \int d^2x\,\left(
 p\eta^{\mu\nu} \eta_{ij} e^i_\mu e^j_\nu
 + q\eta^{\mu\nu} \omega_\mu \omega_\nu\right)$
 has been used with $p$, $q$ being constant parameters,
 and $\eta^{\mu\nu}=$ diag$(+1,-1)$ is the metric of the
 two-dimensional Minkowski space.

Again, the vacuum functional $Z$ is obtained by substituting the explicit 
solutions for the tree action $S$, eq.~(\ref{2grS}), and the gauge-fixing
functional $X$, eq.~(\ref{2grX}), into the expression eq.~(\ref{Z}). The
symmetry transformations (\ref{exBRST}) will not be
given explicitly; their determination is straightforward but the result is
quite lenghty.

Performing the integration over the variables
$\phi^*_{Aa},\pi^{Aa},{\bar\phi_A}$ and $\lambda^A$, we obtain $Z$ in the
form (\ref{w2Z}) with the gauge-fixed effective action $S_{\rm eff}$,
\begin{eqnarray}
\label{PP}
 S_{\rm eff}\!\!\!&=&\!\!\!\!\int {d^2}x\;e\bigg(\frac{1}{16\alpha}
               R_{\mu\nu}{}^{ij}R^{\mu\nu}{}_{ij}
               -\frac{1}{8\beta}T_{\mu\nu}{}^{i}
               T^{\mu\nu}{}_{i}-\gamma\bigg)
	       \\
            &+&\!\!\!\!\!\int d^2x\bigg\{\!p\eta^{\mu\nu}e_{\mu i}\bigg(
           \varepsilon^{ij}e_{\nu j}B+e^i_\lambda\partial_\nu B^\lambda
               +B^\lambda\partial_\lambda e^i_\nu\bigg)
               \!+\!q\eta^{\mu\nu}\omega_\mu\bigg
                (\omega_\lambda\partial_\nu B^\lambda
               +B^\lambda\partial_\lambda\omega_\nu-\partial_\nu B
               \bigg)\!\bigg\}\nonumber\\
           &+&\!\!\!\!\!\frac{1}{2}\varepsilon_{ab}
              \bigg\{p\eta_{ij}\eta^{\mu\nu}\bigg(\varepsilon^{ik}e_{\nu k}
          C^b+e^i_\lambda\partial_\nu C^{\lambda b}
            +C^{\lambda b}\partial_\lambda e^i_\nu\bigg)\bigg(
          \varepsilon^{jl}e_{\mu l}C^a+e^j_\sigma\partial_\mu
              C^{\sigma a}+C^{\sigma a}\partial_\sigma e^j_\mu\bigg)
              \nonumber\\
       &+&\!\!\!\!p\eta^{\mu\nu}\eta_{ij}e^j_\nu\bigg[
          \bigg(e^i_\mu C^b
              +\varepsilon^{ik}e_{\lambda k}\partial_\mu
          C^{\lambda b}
              +\varepsilon^{ik}C^{\lambda b}\partial_\lambda
          e_{\mu k}\bigg)C^a \nonumber\\
          &+&\!\!\!\!\bigg(\varepsilon^{ik}e_{\lambda k}C^b
              +e^i_\sigma\partial_\lambda C^{\sigma b}+C^{\sigma b}
          \partial_\sigma e^i_\lambda\bigg)\partial_\mu C^{\lambda a}
          -C^{\lambda a}\partial_\lambda\bigg(\varepsilon^{ik}e_{\mu k}C^b
              +e^i_\sigma\partial_\mu C^{\sigma b}+C^{\sigma b}
              \partial_\sigma e^i_\mu\bigg)\bigg]\nonumber\\
           &+&\!\!\!q\eta^{\mu\nu}\bigg(\omega_\lambda\partial_\nu C^{\lambda b}
          \!+\!C^{\lambda b}\partial_\lambda\omega_\nu
          \!-\!\partial_\nu C^b\bigg)\bigg(\omega_\sigma\partial_\mu
          C^{\sigma a}
          \!+\!C^{\sigma a}\partial_\sigma\omega_\mu
          \!-\!\partial_\mu C^a\bigg)\nonumber\\
       &-&\!\!\!\!q\eta^{\mu\nu}\omega_\nu\bigg[C^{\lambda a}\partial_\lambda
              \bigg(\omega_\sigma\partial_\mu C^{\sigma b}
              \!+\!C^{\sigma b}\partial_\sigma\omega_\mu
              \!-\!\partial_\mu C^b\bigg)
\label{2grS2}
       \!\!-\!\!\bigg(\omega_\sigma\partial_\lambda C^{\sigma b}
       \!+\!C^{\sigma b}\partial_\sigma\omega_\lambda
       \!-\!\partial_\lambda C^b
       \bigg)\partial_\mu C^{\lambda a}\bigg]\!\bigg\}.\nonumber
\end{eqnarray}
 The effective action $S_{\rm eff}$ is
 invariant under the following transformations:
\begin{eqnarray}
 \delta e^i_\sigma\!\!&=&\!\!\left(\varepsilon^{ij}e_{\sigma j}C^a+
                     C^{\lambda a}\partial_\lambda e^i_\sigma
                     +e^i_{\lambda}\partial_\sigma C^{\lambda a}\right)\mu_a,
                     \nonumber\\
 \delta\omega_\sigma\!\!&=&\!\!\left(-\partial_\sigma C^a+C^{\lambda a}
                       \partial_\lambda\omega_\sigma
                       +\omega_\lambda\partial_\sigma C^{\lambda a}
                       \right)\mu_a,\nonumber\\
 \delta B\!\!&=&\!\!\hbox{\large$\frac{1}{2}$}
 \Big(C^{\sigma a}\partial_\sigma B
            -B^\sigma\partial_\sigma C^a+
	    \hbox{\large$\frac{1}{6}$}\varepsilon_{bd}
            (C^{\sigma b}\partial_\sigma(C^{\lambda \{a}\partial_\lambda
            C^{d\}})-(C^{\lambda\{a}\partial_\lambda C^{\sigma d\}})
            \partial_\sigma C^b\Big)\mu_a,\nonumber\\
 \delta B^{\sigma}\!\!&=&\!\!\hbox{\large$\frac{1}{2}$}
 \Big(C^{\lambda a}\partial_\lambda B^\sigma
                    -B^\lambda\partial_\lambda C^{\sigma a}
                    +\hbox{\large$\frac{1}{6}$}\varepsilon_{bd}
                    (C^{\lambda b}\partial_\lambda(C^{\rho \{a}\partial_\rho
                    C^{\sigma d\}})
                    -(C^{\rho \{a}\partial_\rho C^{\lambda d\}})
                    \partial_\lambda C^{\sigma b}\Big)\mu_a,\nonumber\\
 \delta C^a\!\!&=&\!\!\Big(\varepsilon^{ab}B-
 \hbox{\large$\frac{1}{2}$}C^{\sigma \{a}
              \partial_\sigma C^{b\}}\Big)\mu_b,\nonumber\\
\label{2grBRST2}
 \delta C^{\sigma a}\!\!&=&\!\!\Big(\varepsilon^{ab}B^\sigma-
 \hbox{\large$\frac{1}{2}$}
                       C^{\lambda\{a}\partial_\lambda C^{\sigma b\}}
                       \Big)\mu_b.
\end{eqnarray}
 which, consequently, realize the transformations of extended BRST 
 symmetry in
 terms of the variables $\phi^A$ of the complete configuration space. 
 Again, the effective action, using the corresponding (anti)BRST operators, 
 can be written in a compact form:
\begin{eqnarray}
 S_{\rm eff} = S_{\rm cl} + \frac{1}{2} \varepsilon_{ab} s^b s^a
 F(e^i_\mu, \omega_\mu),
\end{eqnarray}
which coincides with the effective action in the
$Sp(2)$-covariant approach \cite{PP}. (In comparision with that reference
some misprints have been removed in eqs.~(\ref{PP})).

\section{Conclusion}

 In this paper we have exemplified the method of modified
 triplectic quantization \cite{GGL} on the basis of several gauge theory
 models. Thus, we have considered the model \cite{FT} of non-abelian
 antisymmetric tensor field (Freedman--Townsend model), the model \cite{W2}
 of $W_2-$gravity, and the model \cite{VK} of two-dimensional gravity with
 dynamical torsion. For these models we have found explicit solutions of the
 generating equations that determine the (tree approximation of the) 
 quantum action $W$ and the gauge-fixing functional $X$ in the
 framework of the modified triplectic formalism \cite{GGL}. In the case of
 the 2-dimensional models we did not determine possible anomalies which
 occure if loop correction are taken into account.

 The above solutions are expressed in terms of the variables $\phi^A$,
 $\phi^*_{Aa}$ and $\pi^{Aa}$, $\bar{\phi}_A$ anticanonically conjugated in
 the sense of the extended antibrackets \cite{BM,GGL}, as well as in terms of
 the additional variables $\lambda^A$ that serve to parametrize 
 the gauge-fixing  functional $X$. 
 However, it should be remarked that by the special choice of both the action
 functional and of the gauge fixing functional triplecticity of the formalism
 is reduced, in fact, to the usual case of the $Sp(2)$-covariant
 quantization.  
 
 Using the solutions for $S$ and $X$, we have obtained the vacuum functional
 and explicitly constructed the corresponding transformations \cite{GGL} of
 extended BRST symmetry in terms of the anticanonically conjugated variables.
 Finally, we have obtained manifest $Sp(2)$-symmetric expressions for 
 the effective action
 $S_{\rm eff}$ that result from integrating out the variables
 $\phi^*_{Aa}$, $\pi^{Aa}$, $\bar{\phi}_A$ and $\lambda^A$ in the functional
 integral, and we have constructed the corresponding transformations
 (\ref{exBRST})
 of extended BRST symmetry in terms of the variables $\phi^A$ of the complete
 configuration space. In the case of the irreducible theories \cite{W2,VK},
 the resulting actions $S_{\rm eff}$ have been represented in a compact form,
 expressing the gauge-fixed part in terms of (anti)BRST variations. 
 In any case we finally obtained a  $Sp(2)$-symmetric action which is
invariant under BRST and antiBRST transformations. In the case of
irreducible theories we were able to write down the gauge fixing part in a
very simple manner.
In the case
 of the first-stage reducible FT model the situation occured much more
difficult. Especially the dependence of the corresponding (anti)BRST
transformation for the classical $H$-field on the gauge parameter $\alpha$
deserves further study.

\vspace{0.5cm}

{\Large
{\bf {Acknowledgments}}}

\vspace{0.5cm}
\noindent
The authors benefited from various discussions with D.M. Gitman. 
B.G. grateful
acknowledges support from the German--Brasil exchange programmes DAAD 
and FAPESP,
as well as the warm hospitality of the Institute of Mathematical Physics 
of the University of S\~{a}o Paulo, where this work was completed. 
The work of P.M.L.
and P.Yu.M. was partially supported by the Russian Foundation for 
 Basic Research
(RFBR), project 99-02-16617, and the Russian Ministry of Education 
(Fundamental Sciences Grant, E00-3.3-461). The work of P.M.L. was supported 
by INTAS, grant 99-0590,
as well as by the joint project of RFBR and Deutsche Forschungsgemeinschaft 
(DFG),
99-02-04022. P.M.L. also gratefully acknowledges the hospitality of 
NTZ at the
Center of Advanced Study of Leipzig University.

\vspace{0.8cm}
\noindent
{\bf \Large  Appendix}
\vspace{0.8cm}

\setcounter{equation}{0}
\renewcommand{\theequation}{A.\arabic{equation}}
\noindent

 Here, we prove the representation (\ref{ftZ}) -- (\ref{ftDelta}) for the
 vacuum functional in the Freedman--Townsend model
(\ref{fts}) -- (\ref{ftreduce}).
 To this end, we consider the vacuum functional (\ref{Z}) with $S$ and $X$
 given by eqs.~(\ref{ftS}) and (\ref{ftX}), respectively. Integrating in
 eq.~(\ref{Z}) over $\lambda^{A}$ and $\bar{\phi}_A$ leads to the
replacement
\begin{eqnarray*}
 \bar{A}_{\mu}=\bar{B}_{\mu}=\bar{B}_a=\bar{C}_{ab}=0,\qquad
 \bar{H}_{\mu\nu}=\frac{\alpha}{2}H_{\mu\nu},\qquad
 \bar{C}_{\mu a}=\beta\varepsilon_{ab}C^b_{\mu}.
\end{eqnarray*}
 Integrating over the variables $\pi^{Aa}$ and the antifields
$\phi^*_{Aa}$,
 except $H^*_{\mu\nu a}$ and $C^*_{\mu a|b}$, with the subsequent
replacements
\begin{eqnarray*}
 B^*_{\mu a}=C^*_{a|bc}=0,
\end{eqnarray*}
 leads to
\begin{eqnarray}
\label{Z2}
 Z=\int d\phi\,dH^*_{\mu\nu a}dC^*_{\mu a|b}
 \exp\left\{\frac{i}{\hbar}\tilde{S}(\phi,H^*,C^*)\right\},
\end{eqnarray}
 with
\begin{eqnarray*}
 \tilde{S}&=&S_{\rm cl}+\int d^4x\,\bigg\{H_{\mu\nu a}^\ast({\cal D}^\mu C^{\nu
a}-
 {\cal D}^\nu C^{\mu a})-\varepsilon^{ab}C_{\mu a|b}^\ast B^{\mu}
 -\frac{\alpha}{2}H_{\mu\nu}({\cal D}^\mu B^{\nu}-{\cal D}^\nu
B^{\mu})\nonumber\\
 &+&C_{\mu a|b}^\ast {\cal D}^\mu C^{ab}
 +\beta\varepsilon_{ab}C^b_{\mu}{\cal D}^\mu B^a
 +\frac{1}{2}\varepsilon^{\mu\nu\rho\sigma}(H_{\mu\nu a}^\ast\wedge
 H_{\rho\sigma b}^\ast)C^{ab}
 +\frac{\alpha}{4}\varepsilon^{\mu\nu\rho\sigma}
 H_{\mu\nu a}^\ast(H_{\rho\sigma}\wedge B^a)\bigg\}\nonumber\\
 &+&\int d^4x\,\bigg\{\frac{1}{\alpha}
 H^*_{\mu\nu a}H^{*\mu\nu}_b\varepsilon^{ab}
 +\frac{1}{2\beta}C^*_{\mu
a|b}C^{*\mu}_{c|d}\varepsilon^{ac}\varepsilon^{bd}
 \bigg\},
\end{eqnarray*}
 where $S_{\rm cl}$ is the classical action (\ref{fts}) of the Freedman--Townsend
model.

 Let us represent $Z$ in eq.~(\ref{Z2}) as
\begin{equation}
\label{Z3}
 Z=\int d\phi\,\exp\left\{\frac{i}{\hbar}S_{\rm eff}(\phi)\right\},
\end{equation}
where
\begin{eqnarray}
\label{Seff}
 \exp\left\{\frac{i}{\hbar}S_{\rm eff}\right\}
 =\int dH^*_{\mu\nu a}dC^*_{\mu a|b}
 \exp\left\{\frac{i}{\hbar}\tilde{S}(\phi,H^*,C^*)\right\}.
\end{eqnarray}
 It is convenient to rewrite eq.~(\ref{Seff}) in the form
\begin{eqnarray}
\label{Srepr}
 \exp\left\{\frac{i}{\hbar}S_{\rm eff}\right\}
 =C\cdot H\exp\left\{\frac{i}{\hbar}\left[S_0+\int d^4x
 \left(-\alpha B_{\mu\nu}{\cal D}^\mu B^\nu+\beta\varepsilon_{ab}C^b_{\mu}
 {\cal D}^{\mu}B^a\right)\right]\right\}\,,
\end{eqnarray}
 with $C$ and $H$ being integrals over $C^*_{\mu a|b}$ and
 $H^*_{\mu\nu a}$, respectively,
\begin{eqnarray}
\label{BC}
 C&=&\int dC^*_{\mu a|b}\exp\left\{\frac{i}{\hbar}\int d^4x\left(
 C^*_{\mu a|b}\left({\cal D}^{\mu}C^{ab}-\varepsilon^{ab}B^{\mu}\right)
 +\frac{1}{2\beta}C^*_{\mu a|b}C^{*\mu}_{c|d}
 \varepsilon^{ac}\varepsilon^{bd}
 \right)\right\},\nonumber\\
 H&=&\int dH^*_{\mu\nu a}\exp\bigg\{\frac{i}{\hbar}\int
d^4x\bigg(H^*_{\mu\nu a}
 G^{\mu\nu a}+\frac{1}{\alpha}H^*_{\mu\nu a}H^{*\mu\nu}_b\varepsilon^{ab}
 +\frac{1}{2}\varepsilon^{\mu\nu\rho\sigma}
 (H^*_{\mu\nu a}\wedge H^*_{\rho\sigma b})C^{ab}\bigg)\bigg\},\nonumber\\
\end{eqnarray}
 where the object $G^a_{\mu\nu}$ is given by eq.~(\ref{KG}).

 The integral $C$ in eq.~(\ref{BC}) is Gaussian and can be easily
 calculated,
\begin{equation}
 \label{C2}
 C=\exp\left\{-\frac{i}{\hbar}\int d^4x\left(\beta B_{\mu}B^{\mu}
 +\frac{\beta}{2}\varepsilon_{ab}\varepsilon_{cd}({\cal D}_{\mu}C^{ac})
 ({\cal D}^{\mu}C^{bd})\right)\right\}.
\end{equation}
 In order to calculate $H$ in eq.~(\ref{BC}), we use the decompositions
\begin{eqnarray}
 \label{decomp}
  H^*_{\mu\nu a}&=&\left(H^*_{oi a},H ^*_{ia}\right),\;\;
  H^*_{ia}\equiv\frac{1}{2}\varepsilon_{oijk}H^{*jk}_a,\nonumber\\
  G^a_{\mu\nu}&=&\left(G^a_{oi},G ^a_{i}\right),\;\;
  G^a_{i}\equiv\frac{1}{2}\varepsilon_{oijk}G^{jka}
\end{eqnarray}
 and the following relations:
\begin{eqnarray*}
 &&H^*_{\mu\nu a}G^{\mu\nu a}=2(H^*_{oi a}G^{oi a}-H^*_{i a}G^{i
a}),\nonumber\\
 &&\varepsilon^{ab}H^*_{\mu\nu a}H^{*\mu\nu}_b=2\varepsilon^{ab}(H^*_{oi
a}H^{*oi}_b
 -H^*_{ia}B^{*i}_b),\nonumber\\
 &&\frac{1}{2}\varepsilon^{\mu\nu\rho\sigma}
 (H^*_{\mu\nu a}\wedge H^*_{\rho\sigma b})C^{ab}=4(H^*_{oi a}\wedge
H^{*i}_b)C^{ab}.
\end{eqnarray*}
 Integrating out the components $H^*_{ia}$ allows to represent $H$
 in eq.~(\ref{BC}) as
\begin{eqnarray}
\label{B2}
 H&=&\int dh^*_{ia}\exp\bigg\{\frac{i}{\hbar}\int
d^4x\bigg(\frac{1}{2\alpha}
 h^*_{ia}h^{*i}_b\varepsilon^{ab}+h^*_{ia}G^{oi a}\nonumber\\
 &-&\frac{\alpha}{2}\varepsilon_{ab}(G^a_i
 -h^*_{ic}\wedge C^{ac})(G^{ib}-h^{*i}_d\wedge C^{bd})\bigg)\bigg\},
\end{eqnarray}
 where, for brevity, we denote $h^*_{ia}\equiv 2H^*_{oi a}$.

 Let us further rewrite $H$ in eq.~(\ref{B2}) as
\begin{equation}
\label{B3}
 H=\int dh^*_{ia}\exp\left\{\frac{i}{\hbar}\int d^4x\left(
 -\frac{\alpha}{2}\varepsilon_{ab}G^a_iG^{ib}+h^*_{ia}\bar{G}^{ia}
 +\frac{1}{2\alpha}h^*_{ia}{\cal M}^{ab}h^{*i}_b\right)\right\},
\end{equation}
 where
\begin{equation}
\label{Gbar}
 \bar{G}^a_i\equiv G^a_{0i}+\alpha {\cal C}^a_b G^b_i,
\end{equation}
 and the objects ${\cal C}^a_b$, ${\cal M}^{ab}$ are given by
 eqs.~(\ref{M^ab}), (\ref{CB}). From the definitions (\ref{M^ab})
 and (\ref{CB}) follow the algebraic properties
\[
 ({\cal C}^a_bF)G=-F({\cal C}^a_bG),\;\;
 \varepsilon^{ac}{\cal C}^b_c=-{\cal C}^a_c\varepsilon^{cb},\;\;
 {\cal C}^c_a\varepsilon_{cb}=-\varepsilon_{ac}{\cal C}^c_b,
\]
\begin{eqnarray}
 \label{alg}
 {\cal M}^{ac}{\cal C}_c^b=-{\cal C}^a_c{\cal M}^{cb},\;\;
 {\cal M}^{-1}_{ac}{\cal C}_b^c=-{\cal C}^c_a{\cal M}^{-1}_{cb},
\end{eqnarray}
 where $F\equiv F^p$, $G\equiv G^p$ are arbitrary objects carrying the
 gauge indices $p$, and the matrix ${\cal M}^{-1}_{ab}$ is the inverse
 of ${\cal M}^{ab}$ as defined in eq.~(\ref{M^ab}).

 Integrating in eq.~(\ref{B3}) over $h^*_{ia}$ leads to
\begin{equation}
 \label{B4}
 H=\exp\left\{\frac{i}{\hbar}Y\right\}
 \int dh^*_{ia}\exp\left\{\frac{i}{2\alpha\hbar}\int d^4x\; h^*_{ia}
 {\cal M}^{ab}h^{*i}_b\right\},
\end{equation}
 where
\begin{eqnarray}
\label{Y}
  Y=\int d^4x\left\{\frac{\alpha}{2}\bar{G}^a_i{\cal
M}^{-1}_{ab}\bar{G}^{ib}
  -\frac{\alpha}{2}G^a_iG^{ib}\varepsilon_{ab}\right\}
  \equiv\int d^4x\,{\cal Y}.
\end{eqnarray}
 By virtue of the definition (\ref{Gbar}), the integrand ${\cal Y}$ in
 eq.~(\ref{Y}) takes the form
\begin{eqnarray}
\label{Y2}
 {\cal Y}&=&\frac{\alpha}{2}G^a_{0i}{\cal M}^{-1}_{ab}G^{oib}
 -\frac{\alpha}{2}\varepsilon_{ab}G^a_{i}G^{ib}
 +\frac{\alpha^2}{2}G^a_{0i}{\cal M}^{-1}_{ab}{\cal C}^b_c
G^{ic}\nonumber\\
 &&+\frac{\alpha^2}{2}{\cal C}^a_c G^c_i{\cal M}^{-1}_{ab}G^{oib}
 +\frac{\alpha^3}{2}{\cal C}^a_c G^c_i{\cal M}^{-1}_{ab}{\cal C}^b_d
G^{id}.
\end{eqnarray}
  From the algebraic properties (\ref{alg}) it follows that
\begin{eqnarray}
\label{sum1}
 {\cal C}^a_c G^c_i{\cal M}^{-1}_{ab}G^{oib}
 =-G^c_i{\cal C}^a_c{\cal M}^{-1}_{ab}G^{oib}
 =G^c_i{\cal M}^{-1}_{ca}{\cal C}^a_bG^{oib}.
\end{eqnarray}
 Similarly we have
\begin{eqnarray}
\label{sum2}
 {\cal C}^a_c G^c_i{\cal M}^{-1}_{ab}{\cal C}^b_d G^{id}
 =-G^c_i{\cal C}^a_c{\cal M}^{-1}_{ab}{\cal C}^b_d G^{id}
 =G^c_i{\cal M}^{-1}_{ca}{\cal C}^a_b{\cal C}^b_d G^{id}.
\end{eqnarray}
 Note further that the combination ${\cal C}^a_b{\cal C}^b_d$
 in eq.~(\ref{sum2}) can be represented as
\begin{equation}
\label{mult}
 {\cal C}^a_b {\cal C}^b_d
 =\frac{1}{\alpha^2}({\cal M}^{ab}-\varepsilon^{ab})\varepsilon_{bd}.
\end{equation}
 Indeed, by virtue of the property
 ${\cal C}^c_a\varepsilon_{cb}=-\varepsilon_{ac}{\cal C}^c_b$ in
 eq.~(\ref{alg}), we have
\begin{equation}
\label{indeed}
 {\cal C}^a_c {\cal C}^c_b
 ={\cal C}^a_c {\cal C}^d_b\delta^c_d
 ={\cal C}^a_c {\cal C}^d_b\varepsilon^{cp}\varepsilon_{pd}
 =\varepsilon^{cp}{\cal C}^a_c {\cal C}^d_b\varepsilon_{pd}
 =-\varepsilon^{cp}{\cal C}^a_c{\cal C}^d_p\varepsilon_{db},
\end{equation}
 which is equivalent to (\ref{mult}) with allowance for
 the definition
 ${\cal M}^{ab}\equiv\varepsilon^{ab}-\alpha^2\varepsilon^{cd}{\cal C}^a_c
 {\cal C}^b_d$.
    
 Using eqs.~(\ref{sum1}) -- (\ref{mult}) and the fact that
 ${\cal M}^{ac}{\cal M}^{-1}_{cb}=\delta^a_b$, we find that eq.~(\ref{Y2})
 becomes
\begin{eqnarray*}
 {\cal Y}=\frac{\alpha}{2}\left(G^a_{0i}{\cal M}^{-1}_{ab}G^{oib}
 -G^a_{i}{\cal M}^{-1}_{ab}G^{ib}\right)
 +\frac{\alpha^2}{2}\left(
 G^a_{0i}{\cal M}^{-1}_{ab}{\cal C}^b_cG^{ic}
 +G^a_{i}{\cal M}^{-1}_{ab}{\cal C}^b_cG^{0ic}\right),
\end{eqnarray*}
 which can be represented in a manifestly covariant form
\begin{eqnarray}
\label{Y3}
 {\cal Y}=\frac{\alpha}{4}G^a_{\mu\nu}{\cal M}^{-1}_{ab}G^{\mu\nu b}
 +\frac{\alpha^2}{8}\varepsilon^{\mu\nu\rho\sigma}G^a_{\mu\nu}
 {\cal M}^{-1}_{ab}{\cal C}^b_cG^c_{\rho\sigma}.
\end{eqnarray}
 From eq.~(\ref{Y3}) it follows that $H$ in eq.~(\ref{B4}) is given by
\begin{equation}
\label{B5}
 H=\Delta\exp\left\{\frac{\alpha i}{4\hbar}\int d^4x\,
 G^a_{\mu\nu}{\cal M}^{-1}_{ab}\left(
 \frac{1}{2}\delta^b_c(\eta^{\mu\rho}\eta^{\nu\sigma}-\eta^{\mu\sigma}
 \eta^{\nu\rho})+\frac{\alpha}{2}\varepsilon^{\mu\nu\rho\sigma}
 {\cal C}^b_c\right)G^c_{\rho\sigma}\right\},
\end{equation}
 where $\Delta$ denotes the integral over $h^*_{ia}$ in eq.~(\ref{B4}) and
 coincides with the corresponding object introduced in eq.~(\ref{ftDelta}).

 Collecting eqs.~(\ref{Srepr}), (\ref{C2}) and (\ref{B5}), we conclude that
 the vacuum functional (\ref{Z3}) for the Freedman--Townsend model
 coincides with the representation (\ref{ftZ}) -- (\ref{ftDelta}).

\end{document}